\documentclass[a4paper]{article}
\usepackage{amssymb}
\newcommand{\bea}{\begin{eqnarray}}
\newcommand{\eea}{\end{eqnarray}}
\newcommand{\ba}{\begin{array}}
\newcommand{\ea}{\end{array}}
\newcommand{\edc}{\end{document}}
\newcommand{\bc}{\begin{center}}
\newcommand{\ec}{\end{center}}
\newcommand{\be}{\begin{equation}}
\newcommand{\ee}{\end{equation}}

\newcommand{\dsf}{\displaystyle\frac}
\def\s{\sigma}

\def\l{\lambda}
\def\z{\eta}

\def\Q{\mathbb{Q}}
\def\Z{\mathbb{Z}}
\def\a{\alpha}

\def\O{\Omega}

\def\h{\mathbf{h}}
\def\z{\mathbf{s}}

\def\b{\beta}
\def\m{\mu}

\begin{document}
\begin{center}
{\bf ON UNIQUENESS OF GIBBS MEASURES FOR $P$-ADIC NONHOMOGENEOUS $\l$-MODEL ON THE CAYLEY TREE}\\[1.4cm]
MUROD ~ KHAMRAEV\\[1mm]
\small {\it Institute of Mathematics, 29,  F.Hodjaev str.,
Tashkent, 700143,
Uzbekistan\\
E-mail: dagnirmor@wiut.uz}\\[4mm]
FARRUH  ~ MUKHAMEDOV\\[1mm]
\small {\it
Department of Mechanics and Mathematics, \\
National University of Uzbekistan,\\
Vuzgorodok, 700095,  Tashkent, Uzbekistan\\
E-Mail: far75m@yandex.ru}\\[4mm]
UTKIR ~ ROZIKOV\\[1mm] \small {\it Institute of Mathematics,
29,  F.Hodjaev str., Tashkent, 700143,
Uzbekistan\\
E-mail: rozikovu@yandex.ru}\\[1.5cm]
\end{center}
\begin{abstract}
We consider a nearest-neighbor $p$-adic $\l$-model with spin
values $\pm 1$ on a Cayley tree of order $k\geq 1$. We prove for
the model there is no phase transition and  as well as the unique
$p$-adic Gibbs measure is bounded if and only if  $p\geq 3$. If
$p=2$ then we find a condition which guarantees nonexistence of a
phase transition. Besides, the results are applied to the $p$-adic
Ising model and we show that
for the model there is a unique $p$-adic Gibbs measure. 
\end{abstract}


\section{Introduction}

Interest in the physics of non-Archimedean quantum models
\cite{ADFV},\cite{FW},\cite{VVZ},\cite{MP},\cite{V2} is based on
the idea that the structure of space-time for very short distances
might conveniently be described in terms of non-Archimedean
numbers. One of the ways to describe this violation of the
Archimedean axiom, is the using $p$-adic analysis (see \cite{Sh}).
Numerous applications of this analysis to mathematical physics
have been proposed in papers
\cite{ADFV},\cite{K1},\cite{K2},\cite{MP},\cite{V2}. It is known
\cite{K2} that a number of $p$-adic models in physics cannot be
described using ordinary Kolmogorov's probability theory
\cite{Ko}. New probability models - $p$-adic probability models
were investigated in \cite{K2},\cite{K3}. In \cite{KL1},\cite{KL2}
the theory of stochastic processes with values in $p$-adic and
more general non-Archimedean fields having probability
distributions with non-Archimedean values has been developed. The
non-Archimedean analog of the Kolmogorov theorem was proved, that
allows to construct wide classes of stochastic processes by using
finite dimensional probability distributions. This gives a
possibility to develop the theory of statistical mechanics in the
context of the $p$-adic theory, since it lies on the base of the
theory of probability and stochastic processes.

One of the central problems of the theory of statistical mechanics
is the study of infinite-volume Gibbs measures corresponding to a
given Hamiltonian. However, a complete analysis of the set of
Gibbs measures for a specific Hamiltonian is often a difficult
problem. This problem includes the study of phase transitions
problem. Recall that for a given Hamiltonian there is a phase
transition if there exist at least two distinct Gibbs measures.

In this paper we develop $p$-adic probability theory approaches to
study some of statistical mechanics models on a Cayley tree in the
field of $p$-adic numbers. In \cite{GMR} we have proved the
existence of the phase  transition for the homogeneous $p$-adic
Potts model with $q\geq 3$ spin variables on the set of integers
$\mathbb{Z}$. The present paper deals with nonhomogeneous $p$-adic
$\l$-model on the Cayley tree of order $k$, $k\geq 1$. The aim of
this paper is to show the uniqueness of Gibbs measures for the
considered model. In the last section we will apply the obtained
results to the $p$-adic Ising model. Here it should be noted that
the existence of a phase transition for the Ising model (real
case) on the Cayley tree of order $k\geq 2$ was established by
Katsura and Takisawa \cite{KT}.

\section{Definitions and preliminary results}

\subsection{$p$-adic numbers and measures}

Throughout the paper $p$ will be a fixed prime number. Every
rational number $x\neq 0$ can be represented in the form
$x=p^r\dsf{n}{m}$, where $r,n\in\mathbb{Z}$, $m$ is a positive
integer, $(p,n)=1$, $(p,m)=1$. The $p$-adic norm of $x$ is given
by
$$
|x|_p=\left\{ \ba{ll}
p^{-r} & \ \textrm{ for $x\neq 0$}\\
0 &\ \textrm{ for $x=0$}.\\
\ea \right.
$$

It satisfies the following the strong triangle inequality
$$
|x+y|_p\leq\max\{|x|_p,|y|_p\},
$$
this is a non-Archimedean norm.

The completion of the field of rational numbers $\Q$  with respect
to $p$-adic norm is called $p$-adic field and it is denoted by
$\Q_p$.

Let $B(a,r)=\{x\in \Q_p : |x-a|_p< r\}$, where $a\in \Q_p$, $r>0$.
The $p$-adic logarithm is defined by the series
$$
\log_p(x)=\log_p(1+(x-1))=\sum_{n=1}^{\infty}(-1)^{n+1}\dsf{(x-1)^n}{n},
$$
which converges for $x\in B(1,1)$; the $p$-adic exponential is
defined by
$$
\exp_p(x)=\sum_{n=1}^{\infty}\dsf{x^n}{n!},
$$
which converges for $x\in B(0,p^{-1/(p-1)})$.

{\bf Lemma 2.1.}\cite{Kl},\cite{VVZ} {\it Let $x\in
B(0,p^{-1/(p-1)})$ then we have $$ |\exp_p(x)|_p=1,\ \ \
|\exp_p(x)-1|_p=|x|_p<1, \ \ |\log_p(1+x)|_p=|x|_p<p^{-1/(p-1)} $$
and $$ \log_p(\exp_p(x))=x, \ \ \exp_p(\log_p(1+x))=1+x. $$ }

Let $(X,{\cal B})$ be a measurable space, where ${\cal B}$ is an
algebra of subsets $X$. A function $\m:{\cal B}\to \Q_p$ is said
to be a {\it $p$-adic measure} if
$$
\mu(\bigcup_{j=1}^{n} A_j)=\sum_{j=1}^{n}\mu(A_j),
$$
for any $A_1,...,A_n\subset{\cal B}$ such that $A_i\cap
A_j=\emptyset$ ($i\neq j$). A $p$-adic measure is called a {\it
probability measure} if $\mu(X)=1$. A $p$-adic probability measure
$\m$ is said to be {\it bounded} if $\sup\{|\m(A)|_p : A\in {\cal
B}\}<\infty $.

For more details about $p$-adic measures we refer the reader to
\cite{K2},\cite{K3}

\subsection{Cayley tree}

A Cayley tree $\Gamma^k$ of order $ k\geq 1 $ is an infinite tree,
i.e., a graph without cycles in which each vertex lies on $ k+1 $
edges. Let $\Gamma^k=(V, L),$ where $V$ is the set of vertices of
$ \Gamma^k$, $L$ is the set of edges of $ \Gamma^k$. The vertices
$x$ and $y$ are called {\it nearest neighbors} and they are
denoted by $l=<x,y>$ if there exists an edge connecting them. A
collection of pairs $<x,x_1>,...,<x_{d-1},y>$ is called a {\it
path} from the point $x$ to the point $y$. The distance $d(x,y),
x,y\in V$, on the Cayley tree, is the length of the shortest path
from $x$ to $y$.

Let $x_0\in V$ be fixed.  We set
$$ W_n=\{x\in V| d(x,x^0)=n\}, $$
$$ V_n=\cup_{m=1}^n W_m=\{x\in V| d(x,x^0)\leq n\}, $$
$$ L_n=\{l=<x,y>\in L | x,y\in V_n\}. $$

Denote
$$
S(x)=\{y\in W_{n+1} :  d(x,y)=1 \} \ \ x\in W_n,
$$
this set is called the set of {\it direct successors} of $x$.
Observe that any vertex $x\neq x^0$ has $k$ direct successors and
$x^0$ has $k+1$.

\subsection{$p$-adic $\l$-model}

We consider a $p$-adic $\l$-model, where the spin takes values in
the set $\Phi=\{-1,1\}\subset \Q_p$ and is assigned to the
vertices of the tree. A configuration $\s$ on $V$ is then defined
as a function $x\in V\to\s(x)\in\Phi$; in a similar fashion one
defines a configuration $\s_n$ and $\s^{(n)}$ on $V_n$ and $W_n$
respectively. The set of all configurations on $V$ (resp. $V_n$,
$W_n$) coincides with $\Omega=\Phi^{V}$ (resp.
$\Omega_{V_n}=\Phi^{V_n},\ \ \Omega_{W_n}=\Phi^{W_n}$). One can
see that $\O_{V_n}=\O_{V_{n-1}}\times\O_{W_n}$. Using this, for
given configurations $\s_{n-1}\in\O_{V_{n-1}}$ and
$\s^{(n)}\in\O_{W_{n}}$ we define their concatenations  by the
formula
$$
\s_{n-1}\vee\s^{(n)}=\bigg\{\{\s_n(x),x\in
V_{n-1}\},\{\s^{(n)}(y),y\in W_n\}\bigg\}.
$$
It is clear that $\s_{n-1}\vee\s^{(n)}\in \O_{V_n}$. Let functions
$\lambda_{x,y}:(u,v)\in \Phi\times \Phi \to \lambda_{x,y}(u,v)\in
\Q_p.$ be given  for each pairs of neighboring vertices $x,y$. The
Hamiltonian $H_n:\Omega_{V_n}\to\Q_p$ of the $p$-adic
inhomogeneous $\l$-model has the form
$$
H_n(\s_n)=\sum_{<x,y>\in L_n}\l_{x,y}(\s_n(x),\s_n(y)) , \ \ n\in
\mathbb{N} \eqno(2.1)
$$
where the sum is taken over all pairs of neighboring vertices
$<x,y>$ and $\s_n\in\O_{V_n}$.

We say that (2.1) is {\it homogeneous $\l$-model} if all functions
$\l_{xy}(u,v)$ do not depend on $x,y$ and in this case we put
$\l(u,v):=\l_{x,y}(u,v), \ \ \forall <x,y>\in L.$

We note that $\l$-model of this type were firstly considered in
\cite{R}.

Before giving a construction of a special class of Gibbs measures
for $p$-adic  $\l$-model on the Cayley tree we need the following

{\bf Lemma 2.2.} {\it Let $h_x, x\in V$ be a $\Q_p$-valued
function such that $h_x\in B(0,p^{-1/(p-1)})$ for all $x\in V$ and
$|\l_{x,y}(u,v)|_p<p^{-1/(p-1)}$ for all $u,v\in\Phi$ and
$<x,y>\in L$. Then the relation
$$
H_n(\s)+\sum_{x\in W_n}h_x\s(x)\in B(0,p^{-1/(p-1)})
$$
is valid for any $n\in \mathbb{N}$.}

The proof easily follows from the strong triangle inequality for
the norm $|\cdot|_p$.

Let $\mathbf{h}:x\in V\to h_x\in\Q_p$ be a function of $x\in V$
such that $|h_x|_p<p^{-1/(p-1)}$ for all $x\in V$. Given
$n=1,2,...$, consider a $p$-adic probability measure $\m^{(n)}_\h$
on $\O_{V_n}$ defined by
$$
\mu^{(n)}_\h(\s_n)=Z^{-1}_{n}\exp_p\{H_n(\s_n)+\sum_{x\in
W_n}h_x\s(x)\}, \eqno(2.2)
$$
Here, as before, $\s_n:x\in V_n\to\s_n(x)$ and $Z_n$ is the
corresponding partition function:
$$
Z_n=\sum_{\tilde\s_n\in\Omega_{V_n}}\exp_p\{H(\tilde\s_n)+\sum_{x\in
W_n}h_x\tilde\s(x)\}.
$$

Note that according to Lemma 2.2 the measures $\m^{(n)}_\h$ exist.

The compatibility conditions for $\m^{(n)}_\h(\s_n), n\geq 1$ are
given by the equality
$$
\sum_{\s^{(n)}\in\O_{W_n}}\m^{(n)}_\h(\s_{n-1}\vee\s^{(n)})=\m^{(n-1)}_\h(\s_{n-1}),
\eqno(2.3)
$$
where $\s_{n-1}\in\O_{V_{n-1}}$.

We note that an analog of the Kolmogorov extension theorem for
distributions can be proved for $p$-adic distributions given by
(2.2) (see \cite{KL2}). Then according to the Kolmogorov theorem
there exists a unique $p$-adic measure $\m_\h$ on $\O=\Phi^V$ such
that for every $n=1,2,...$ and $\s_n\in\Phi^{V_n}$ the equality
holds
$$
\m_\h\bigg(\{\s|_{V_n}=\s_n\}\bigg)=\m^{(n)}_\h(\s_n),
$$
which will be called a {\it $p$-adic Gibbs measure} for the
considered $\l$-model. It is clear that the measure $\m_\h$
depends on the function $h_x$. By ${\cal S}_\l$ we denote the set
of all $p$-adic Gibbs measures associated with functions
$\h=(h_x,\ x\in V)$. If $|{\cal S}_\l|\geq 2$, then we say that
for this model there exists {\it a phase transition}, otherwise,
we say there is {\it no phase transition} ( here $|A|$ means the
cardinality of a set $A$). In other words,  the phase transition
means that there are two different functions $\h=(h_x,\ x\in V)$
and $\z=(s_x,\ x\in V)$ for which there exists two $\m_\h$ and
$\m_\z$ $p$-adic Gibbs measures on $\O$, respectively.

The following statement describes conditions on $h_x$ guaranteeing
that the compatibility conditions are satisfied by measures
$\m^{(n)}_\h(\s_n)$.

{\bf Theorem 2.3.} {\it The measures $\m^{(n)}_\h(\s_n), \
n=1,2,...$ satisfy the compatibility condition (2.3) if and only
if for any $x\in V$ the following equation holds:
$$
h_x=\sum_{y\in S(x)}F_{x,y}(h_y;\l) \eqno(2.4)
$$
where $S(x)$ is the set of all direct successors of $x\in V$ and
$$
F_{x,y}(h,\l)=\dsf{1}{2}\log_p\left(\dsf{\exp_p(\l_{x,y}(1,1))\exp_p(2h)+\exp_p(\l_{x,y}(1,-1))}
{\exp_p(\l_{x,y}(-1,1))\exp_p(2h)+\exp_p(\l_{x,y}(-1,-1))}\right).
$$}

{\bf Proof.} Using (2.2), it is easy to see that (2.3) and (2.4)
are equivalent.(cf. \cite{R}).\\[3mm]

Observe that according to this Theorem the problem of describing
of ${\cal S}$ is reduced to that of solutions of functional
equation (2.4).

\section{Uniqueness of Gibbs measure for the $p$-adic
$\l$-model}

In this section we will show that the phase transition does not
occur for the $p$-adic $\l$-model.

Set
$$
\Xi=\{\h=(h_x,x\in V) : \ h_x \ \textrm{satisfies the equation
(2.4)} \}.
$$

According to Theorem 2.3 the description of ${\cal S}_\l$ is
reduced to the description of elements of the set $\Xi$.

Set
$$
\left. \ba{ll}
 a_{xy}=\exp_p(\l_{xy}(1,1)), \ \ \ \ b_{xy}=\exp_p(\l_{xy}(1,-1)), \\[3mm]
c_{xy}=\exp_p(\l_{xy}(-1,1)), \ \ \ \ d_{xy}=\exp_p(\l_{xy}(-1,-1)).\\[2mm]
\ea \right\} \eqno(3.1)
$$

The following Theorem is the main result of the paper.

 {\bf Theorem 3.1.} {\it Assume that one of the following conditions is
satisfied:}
\begin{enumerate}
   \item[(i)] {\it $p\geq 3$ and $|\l_{x,y}(u,v)|_p\leq \dsf{1}{p}$
for all $<x,y>\in L$, $u,v\in\Phi$;}
   \item[(ii)]{\it $p=2$ and
   $$
|\l_{xy}(1,1)+\l_{xy}(-1,-1)-\l_{xy}(1,-1)-\l_{xy}(-1,1)|_2\leq
\dsf{1}{2^3}\eqno(3.2)
$$ for all $<x,y>\in
L$,$u,v\in\Phi$;}
\end{enumerate}
{\it  Then for the $p$-adic $\l$-model (2.1) on the Cayley tree of
order $k$ there is no phase transition for any prime $p$, i.e.
$|{\cal S}_\l|\leq 1$.}

{\bf Proof.} If $\Xi=\emptyset$, then nothing to prove. So, assume
that $\Xi\neq\emptyset$. To prove Theorem it is enough to show
that any two elements of $\Xi$ coincide with each other. In order
to do this it is enough to show that for arbitrary $\varepsilon>0$
and every $\h=(h_x,x\in V), \z=(s_x,x\in V)\in \Xi$ and $x\in V$
the inequality $|h_x-s_x|_p<\varepsilon$ is valid.

Take $\h=(h_x,x\in V), \z=(s_x,x\in V)\in \Xi$. Let $x\in V$ be an
arbitrary vertex. Using (3.1) and Lemma 2.1, consider the
difference \bea |h_x-s_x|_p=\bigg|\sum_{y\in S(x)}F_{xy}(h_y;\l)-
\sum_{y\in
S(x)}F_{xy}(s_y;\l)\bigg|_p\leq\nonumber\\
\leq\max_{y\in
S(x)}\bigg|F_{xy}(h_y;\l)-F_{xy}(s_y;\l)\bigg|_p=\nonumber\\
=\max_{y\in
S(x)}\dsf{1}{|2|_p}\bigg|\log_p\bigg[\dsf{a_{xy}\exp_p(2h_y)+b_{xy}}{c_{xy}\exp_p(2h_y)+d_{xy}}\cdot
\dsf{a_{xy}\exp_p(2s_y)+b_{xy}}{c_{xy}\exp_p(2s_y)+d_{xy}}-1+1\bigg]\bigg|_p=\nonumber
\\
=\max_{y\in
S(x)}\dsf{1}{|2|_p}\bigg|\dsf{(a_{xy}d_{xy}-b_{xy}c_{xy})(\exp_p(2h_y)-\exp_p(2s_y))}
{(c_{xy}\exp_p(2h_y)+d_{xy})(a_{xy}\exp_p(2s_y)+b_{xy})}\bigg|_p.\nonumber
\eea

We have
\bea
|a_{xy}d_{xy}-b_{xy}c_{xy}|_p=|\exp_p(\l_{xy}(1,1)+\l_{xy}(-1,-1)-\l_{xy}(1,-1)-\l_{xy}(-1.1))-1|_p\leq\nonumber\\
\leq|\l_{xy}(1,1)+\l_{xy}(-1,-1)-\l_{xy}(1,-1)-\l_{xy}(-1,1)|_p\leq\dsf{1}{p}.\nonumber\\[3mm]
|c_{xy}\exp_p(2h_y)+d_{xy}|_p=|\exp_p(2h_y+2\l_{xy}(-1,1))-1+\nonumber
\\[4mm]
+\exp_p(\l(-1,-1))-1+2|_p=\left\{
\ba{ll} 1, \ \ \textrm{if} \ \ p\geq 3\\[3mm]
\dsf{1}{2}, \ \ \textrm{if} \ \ p=2\\[2mm]
\ea \right. \nonumber\eea

Likewise we get
\bea |a_{xy}\exp_p(2s_y)+b_{xy}|_p=\left\{
\ba{ll} 1, \ \ \textrm{if} \ \ p\geq 3,\\[3mm]
\dsf{1}{2}, \ \ \textrm{if} \ \ p=2,\\[2mm]
\ea \right.\nonumber \\[5mm]
|\exp_p(2h_y)-\exp_p(2s_y)|_p=\left\{
\ba{ll} |h_y-s_y|_p, \ \ \textrm{if} \ \ p\geq 3,\\[3mm]
\dsf{1}{2}|h_y-s_y|_2, \ \ \textrm{if} \ \ p=2.\\[2mm]
\ea \right.\nonumber \eea

Now consider two distinct cases with respect to $p$.

Let us assume that $p\geq 3$. In this case by means of above
relations we find
$$
|h_x-s_x|_p\leq\dsf{1}{p}\max_{y\in S(x)}|h_y-s_y|_p.\eqno(3.3)
$$

Now suppose that $p=2$ and the condition (3.2) is satisfied, then
in this case by similar way as above we obtain
$$
|a_{xy}d_{xy}-b_{xy}c_{xy}|_2\leq\dsf{1}{2^3}.
$$
This with the above relations imply that
$$
|h_x-s_x|_2\leq\dsf{1}{2}\max_{y\in S(x)}|h_y-s_y|_2.\eqno(3.4)
$$

Let us numerate elements of $S(x)$ as follows
$S(x)=\{x_1,...,x_k\}$, here as before $S(x)$ is the set of direct
successors of $x$.

Let $\varepsilon>0$ be an arbitrary number. Take
$n_0\in\mathbb{N}$ such that $\dsf{1}{p^{n_0}}<\varepsilon$. Using
(3.3) (resp. (3.4)) we have \bea
|h_x-s_x|_p\leq\dsf{1}{p}|h_{x_{i_0}}-s_{x_{i_0}}|_p
\leq \dsf{1}{p^2}|h_{x_{i_0,i_1}}-s_{x_{i_0,i_1}}|_p\leq\cdots\nonumber\\
\leq\dsf{1}{p^{n_0-1}}|h_{x_{i_0,...,i_{n_0-2}}}-s_{x_{i_0,...,i_{n_0-2}}}|_p\leq\dsf{1}{p^{n_0}}<\varepsilon,\nonumber
\eea here $x_{i_0,...,i_n,j}$, $j=\overline{1,k}$ are direct
successors of $x_{i_0,...,i_n}$, where
$$|h_{x_{i_0,...,i_m}}|_p=\max\limits_{1\leq j\leq
k}\{|h_{x_{i_0,...,i_{m-1},j}}|_p\}.$$ This completes the proof.\\

{\bf Remark 3.1.} In the next section  we will show that ${\cal
S}_\l$ is non-empty for the Ising model.

{\bf Remark 3.2.} From the proof of Theorem 3.1 it might seem that
$\Xi$ consists only of the single element $h\equiv 0$. But, in
general, this is not true.  Now we give an example of homogeneous
$\l$-model, for which $\Xi$ does not contain $h\equiv 0$. For the
sake of simplicity consider the case $p=3$.

Put
$$
\l(1,1)=\log_34, \ \ \ \l(1,-1)=\log_310, \ \ \ \l(-1,1)=0, \ \ \
\l(-1,1)=\log_34.
$$
According to Lemma 2.1 we easily find that the condition (i) of
Theorem 3.1 is satisfied. Now we search a solution of the equation
(2.4) of the form $\h=(h_x=h,\ x\in V)$. Then (2.4) reduces to the
following equation
$$
\left(\dsf{4z+10} {z+4}\right)^k=z,\eqno(3.5)
$$
where $z=\exp_3(2h)$. By using the Hensel Lemma (see \cite{Kl}),
one  can prove that the equation  (3.5) has a unique solution,
such that $|z-1|_3\leq 1/3$, which is obviously different from $z=1$.\\

Let us assume that ${\cal S}_\l\neq \emptyset$. Now we are
interested whether the unique Gibbs measure is bounded. The
following theorem answers to the question.

{\bf Theorem 3.2.} {\it The $p$-adic Gibbs measure $\m$
corresponding to the $p$-adic $\l$- model on the Cayley tree of
order $k$ is bounded if and and only if $p\neq 2$.}

{\bf Proof.} It suffices to show that the values of $\m$ on
cylindrical subsets are bounded.

Let us assume that $p\geq 3$ and estimate $|\m^{(n)}(\s_n)|_p$:
\bea |\m^{(n)}(\s_n)|_p=\bigg|\frac{\exp_p\{\tilde H(\s_n)\}}
{\sum\limits_{\tilde\s_n\in \O_{V_n}}\exp_p\{\tilde
H(\s_n)\}}\bigg|_p=\nonumber\\
\frac{1}{\bigg|\sum\limits_{\tilde\s_n\in \O_{V_n}}(\exp_p\{\tilde
H(\s_n)\}-1)+2^{V_n}\bigg|_p}=1\nonumber \eea here we have used
Lemma 2.1, where
$$ \tilde
H(\s_n)=H(\s_n)+\sum_{x\in W_n}h_x\s(x). $$

Put
$$
p^{x,y}_{uv}=\frac{\exp_p(\l_{x,y}(u,v)+uh_x+vh_y)}{\sum\limits_{i,j\in\Phi}
\exp_p(\l_{x,y}(i,j)+ih_x+ih_y)},\eqno(3.6)
$$ where $u,v\in\Phi$, $<x,y>\in L$.

To prove that the measure $\m$ is not bounded at $p=2$ it is
enough to show that its marginal  measure is not bounded. Let
$\pi=\{...,x_{-1},x_0,x_1,...\}$ be an arbitrary infinite path in
$\Gamma^k$. From (2.2) one can see that a marginal  measure
$\m_{\pi}$ on $\Psi^{\pi}$ has the form $$
\m_{\pi}(\omega_n)=p^{x_{-n},x_{-n+1}}_{\omega(x_{-n})}\prod_{m=-n}^{n-1}p^{x_m,x_{m+1}}_{\omega(x_m)\omega(x_{m+1})},
\eqno(3.7) $$ here $\omega_n :\{x_{-n},...,x_0,...,x_n\}\to \Psi$,
i.e. $\omega_n$ is a configuration on
$\{x_{-n},...,x_0,...,x_n\}$, $p^{xy}_i$ is an invariant vector
for the matrix $(p^{xy}_{ij})_{ij\in\Phi}$.

Using again Lemma 2.1 from (3.6) we infer  that
$$ |p^{x,y}_{ij}|_p=\frac{1}{\bigg|\sum\limits_{i,j\in\Phi}(\exp_p(\l_{x,y}(i,j)+ih_x+ih_y)-1)+
4\bigg|_p}\geq 2^2. \eqno(3.8) $$ for all $i,j$, here again we
have used Lemma 2.1. From (3.7) and (3.8) we find that $\m_{\pi}$
is not bounded. The theorem is proved.

\section{Applications to $p$-adic Ising model}

In this section we will show that the phase transition does not
occur for the $p$-adic Ising model.

Recall what is a $p$-adic Ising model. This is a particular case
of $\l$-model and it corresponds to the function:
$$
\l_{x,y}(u,v)=J_{x,y}uv+u\eta_x+v\eta_y,
$$
where $|J_{x,y}|< p^{-1/(p-1)}$, $|\eta_x|_p < p^{-1/(p-1)}$ and
$<x,y>\in L$, $u,v\in\{-1,1\}$.

Let us check the condition (3.2) for the Ising model. Let $p=2$,
then the condition (3.2) can be written as follows $$
|J_{x,y}+\eta_x+\eta_y+J_{xy}-\eta_x-\eta_y+J_{xy}-\eta_x+\eta_y+J_{xy}+\eta_x-\eta_y|_2=
$$
$$
=|4J_{x,y}|_2\leq\dsf{1}{2^4}.\eqno(4.1) $$  Hence (3.2) is
satisfied.

From now consider two particular cases of the model: \bea
\l_{x,y}(u,v)=J_{x,y}uv \ \ \ \ \ \textrm{inhomogeneous} \nonumber \\
\l_{x,y}(u,v)=\l(u,v)=Juv+(u+v)\eta \ \ \ \ \
\textrm{homogeneous},\nonumber \eea here $|J|< p^{-1/(p-1)}$,
$|\eta|_p < p^{-1/(p-1)}$.

It is easy to see that for the inhomogeneous model the equation
(2.4) has a solution $h=(h_x=0,\ x\in V)$, so ${\cal S}_\l$ is not
empty.

Thus, we can formulate the following

{\bf Theorem 4.1.} {\it Let $k\geq 1$, and
$|J_{x,y}|_p<p^{-1/(p-1)}$ for all $<x,y>\in L$. Then for the
nonhomogeneous $p$-adic Ising model on the Cayley tree of order
$k$ there is a unique Gibbs measure.}\\

Now turn to homogeneous case. We want to show that $\Xi\neq
\emptyset$. We will search a solution of (2.4) of the form
$\h=(h_x=h,\ x\in V)$.

Let $h_x=h$ for all $x\in V$. Then (2.4) has the following form
$$
\left(\dsf{\a z+1} {z+\b}\right)^k=z,
$$
where $z=\exp_p(2h)$, $\a=\exp_p(2J+2\eta)$ and
$\b=\exp_p(2J-2\eta)$.

Denote
$$
f(x)=\left(\dsf{\a x+1}{x+\b}\right)^k.
$$

Let $S_1=\{x\in\Q_p : |x|_p=1\}$. Then it is clear that
$f(S_1)\subset S_1$. Using this fact and Lemma 2.1, for every
$x\in S_1$, we find
$$
|f(x)-1|_p=\bigg|\dsf{(\a-1)x+1-\b}{x+\b}\bigg|_p\left|\sum_{m=0}^{k-1}
\bigg(\dsf{\a x+1}{x+\b}\bigg)^m\right|_p\leq \dsf{1}{p}.
\eqno(4.2)
$$

By the same argument as the proof of Theorem 3.1, with using
(4.1), for every $x,y\in S_1$ one gets
$$
|f(x)-f(y)|_p\leq\dsf{1}{p}|x-y|_p\eqno(4.3)
$$
for every $p\geq 2$.

Thus the inequality (4.3) implies that $f$ is a contraction of
$S_1$, hence $f$ has a unique fixed point $\zeta\in S_1$ such that
$|\zeta-1|_p\leq\frac{1}{p}$ (see (4.2)). So combining Theorem 3.1
we have proved the following

{\bf Theorem 4.2.} {\it Let $k\geq 1$, $|J|_p<p^{-1/(p-1)}$ and
$|\eta|_p<p^{-1/(p-1)}$. Then for the homogeneous $p$-adic Ising
model on the Cayley tree of order $k$ there is a unique Gibbs measure.}\\

{\bf Remark 4.1.} It is known \cite{KT},\cite{Ru} that for the
Ising model on the Cayley tree of order $k\geq 2$ over
$\mathbb{R}$ under some condition on parameter $J_{x,y}$ (resp.
$J$ and $\eta$) there is a phase transition. Theorems 4.1 and 4.2
show the difference between real the Ising model and the $p$-adic
one.

{\bf Remark 4.2.} In \cite{GMR} we proved that a phase transition
can occur for $q$-state ($q\geq 3$) homogeneous $p$-adic Potts
model on $\Z$ whenever $q$ is divided by a prime $p$. It is known
\cite{B} that 2-state Potts model reduces to the Ising model. So,
Theorem 4.2 shows that in case  $p=2$  for the Ising model there
is no phase transition.

From Theorem 3.2 we deduce the following

{\bf Corollary 4.3.} {\it The unique $p$-adic Gibbs measure $\m$
corresponding to the nonhomogeneous (resp. homogeneous) $p$-adic
Ising model on the Cayley tree of order $k$ is bounded if and
only if $p\neq 2$.}

For the $p$-adic Ising model we have seen that a phase transition
does not occur, so Corollary 4.3 implies that if $p=2$, then the
$p$-adic Gibbs measure may not be bounded in this setting.\\

{\bf Acknowledgement} F.M. thanks the Italian NATO-CNR (2003) for
providing financial support and II Universita di Roma "Tor
Vergata" for all facilities they provide.  U.R. thanks Institute
des Hautes Etudes Scientifiques (IHES) for supporting the visit to
Bures-sur-Yvette (IHES, France) in September-December 2003. The
authors also acknowledge with gratitude to Professors I. V.
Volovich and A. Yu. Khrennikov for the helpful comments and
discussions.

The authors also express gratitude to a referee for some useful
observations.

\newpage

\end{document}